\documentclass[onecolumn,12pt]{IEEEtran}
\usepackage{amsmath, graphics,amssymb,epsfig,subfigure}
\usepackage{algorithm}
\usepackage{algorithmic}
\usepackage{cite}
\def\BibTeX{{\rm B\kern-.05em{\sc i\kern-.025em b}\kern-.08em
    T\kern-.1667em\lower.7ex\hbox{E}\kern-.125emX}}

\newlength{\capwidth}
\newcommand{\be}{\begin{equation}}

\newcommand{\ee}{\end{equation}}
\newcommand{\bea}{\begin{eqnarray}}
\newcommand{\eea}{\end{eqnarray}}
\newcommand{\bdp}{\begin{displaymath}}
\newcommand{\edp}{\end{displaymath}}
\usepackage{cite}

\begin{document}
\title{Transmission Schemes based on Sum Rate Analysis in Distributed Antenna Systems}
\author{\authorblockN{Heejin Kim, Sang-Rim Lee, Kyoung-Jae Lee,\\ and Inkyu Lee, \emph{Senior Member, IEEE}} \\
\authorblockA{School of Electrical Eng., Korea University, Seoul, Korea\\
    Email: $\{$heejink, sangrim78, kyoungjae, inkyu$\}$@korea.ac.kr} \\
} \maketitle \thispagestyle{empty}
\begin{abstract}
In this paper, we study single cell multi-user downlink distributed antenna systems (DAS)
where antenna ports are geographically separated in a cell.
First, we derive an expression of the ergodic sum rate for the DAS in the presence of pathloss.
Then, we propose a transmission selection scheme 
based on the derived expressions which does not require channel state information at the transmitter.
Utilizing the knowledge of distance information from a user to each distributed antenna (DA) port,
we consider the optimization of
pairings of DA ports and users to maximize the system performance.
Based on the ergodic sum rate expressions,
the proposed scheme chooses the best mode maximizing the ergodic sum rate
among mode candidates.
In our proposed scheme,
the number of mode candidates are greatly reduced
compared to that of ideal mode selection.
In addition,
we analyze the signal to noise ratio cross-over point for different modes
using the sum rate expressions.
Through Monte Carlo simulations, we show the accuracy of our derivations for the ergodic sum rate.
Moreover, simulation results with the pathloss modeling confirm that
the proposed scheme produces the average sum rate
identical to the ideal mode selection with significantly reduced candidates.
\end{abstract}
\section{Introduction}

Wireless communication systems have been evolving to maximize data
rates for satisfying demands on high speed data and multimedia
services. One of key technologies in next generation communication
systems is a multiple-input multiple-output (MIMO) method, which
enables to increase spectral efficiency without requiring additional
power or bandwidth consumption \cite{Foschini:98,Lee:06JSAC,Lee:07A,INKYU:04}. Recently, a
distributed antenna system (DAS) has been introduced as a new
cellular communication structure for expanding coverage and increasing
sum rates by having distributed antenna (DA) ports throughout a cell.
The DAS is regarded as a potential solution for next generation wireless systems because of its power and capacity merit
over conventional cellular systems which have centralized antennas at the center location \cite{Choi:07}.

Utilizing geographically separated antenna ports,
many works have attempted to enhance the system performance of the DAS including the design of antenna locations \cite{Wang:09} \cite{Park:11}.
Most studies were focused on the uplink performance analysis to exploit its structurally simple feature \cite{Clark:01,Dai:06,Rho:02}.
Recently, the analysis for downlink was studied in \cite{Hasegawa:03} and \cite{Xiao:03}
from an information theoretic point of view.
Also, the downlink capacity under a single user scenario was investigated in \cite{Choi:07}
for the cases with and without perfect channel state information (CSI) at the transmitter.
Recently, \cite{Feng:09} and \cite{Feng1:09} presented downlink sum rate analysis and power allocation methods
with large system limit using random matrix theory.
However, it is difficult to understand the properties of practical finite systems
with the results derived under the large system limit.

Unlike conventional MIMO systems,
all DA ports in DAS have different channel characteristics since the signal from users to DA ports experiences independent
large-scale fadings.
Therefore, we can increase the ergodic sum rate of the DAS by coordinating a transmission technique
for DA ports
in the presence of the large-scale fadings.
A representative work of transmission selection in DAS with
a single user was made in \cite{Hu:06} where an adaptive mode
selection method is designed based on the received signal to noise ratio (SNR) from each DA
port and the required quality-of-service (QoS) of the users.
In \cite{Gan:07}, this result was
extended to an uplink DAS scenario with two users and two DA ports,
and an adaptive transmit-receive mode selection scheme was proposed.
However,
these works focused on the condition which satisfies the QoS requirement
and mainly studied the error performance for transmission methods.

In this paper, we investigate a transmission scheme for multi-user multi DA port DAS to maximize the ergodic sum rate.
There are several possible transmission modes depending on the pairings of users and DA ports.
First, utilizing the user distance information,
we derive an expression of the ergodic sum rate for various transmission modes.
Then, we introduce a transmission selection scheme which adaptively determines the best transmission mode among mode candidates
using only the knowledge of the distance from users to DA ports without requiring instantaneous CSI.
Moreover, our proposed scheme significantly reduces the number of mode candidates
without any loss of the ergodic sum rate performance.
In addition, by applying an approximation to the derived ergodic sum rate expressions,
we obtain the SNR cross-over points for different transmission modes,
which provide helpful insights on the transmission mode selection.
In the simulation section, we confirm the accuracy of the derived expressions
by evaluating the cell averaged ergodic sum rate performance.
Also, we show that the proposed scheme yields the performance identical to the
ideal transmission selection case with much reduced candidates.

The remainder of the paper is organized as follows: In Section \ref{sec:sec2},
we present the channel model of DAS with single cell multi-user environments.
Section \ref{sec:Ergodic_SR} derives an ergodic sum rate expression for transmission modes.
In Section \ref{sec:schemes}, we propose transmission mode selection strategies based on the derived expressions
and present analysis of the mode selection methods in Section \ref{sec:Analysis}.
We provide simulation results in Section \ref{sec:simulation}.
Finally, Section \ref{sec:conclusion} concludes this paper.

Throughout this paper, $\mathbb{E}[\cdot]$ represents the expectation operation
and $\text{max}(a,b)$ denotes the maximum of $a$ and $b$.
%

\section{System Model} \label{sec:sec2}
\begin{figure}
\begin{center}
\includegraphics[width=3.5in]{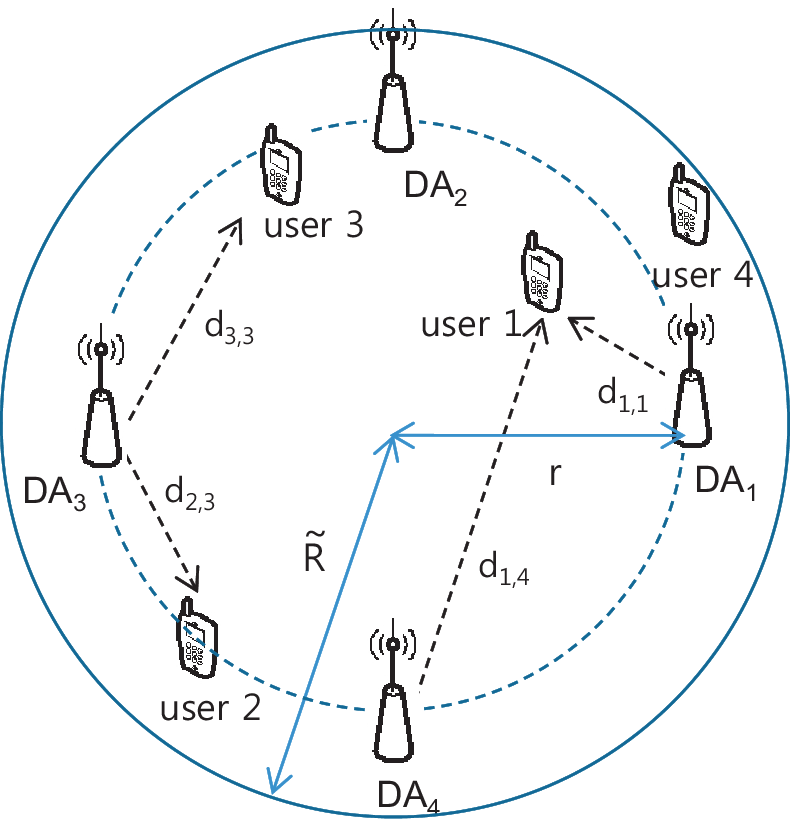}
\end{center}
\caption{Structure of DAS with four users and four distributed antenna ports ($N =K= 4$)}
\label{figure:sys_model}
\end{figure}

We consider single cell downlink DAS environments with $K$ users and $N$ DA ports,
all equipped with a single antenna as shown in Figure \ref{figure:sys_model}.
In our analysis, it is assumed that instantaneous CSI is available only at the receiver side and
each DA port has individual power constraint $P$.
We assume that all DA ports are physically connected with each other via
dedicated channels such as fiber optics and an exclusive RF link.
Moreover, we assume that all DA ports share user data and user distance information,
but do not require the CSI of each user.
The user distance information can be simply obtained by measuring the received signal strength indicator \cite{Wimax} and thus
the amount of feedback is significantly reduced compared to the system which requires the instantaneous CSI.

The received signal for the $i$-th user is written as
\bea
\label{eq:sys1}
y_i=\sum^{N}_{j=1}\sqrt{S_{i,j}P} h_{i,j}x_{j}+ z_{i} ~~~~\text{     for } i=1\text{, }2\text{,..., }K\nonumber
\eea
where $S_{i,j}=d_{i,j}^{-p}$ denotes the propagation pathloss with the pathloss exponent $p$
due to the distance $d_{i,j}$ between the $i$-th user and the $j$-th DA port,
$P$ equals the transmit power,
$h_{i,j}$'s indicate independent and identically distributed
complex Gaussian random variables with unit variance,
$x_j$ stands for the transmitted symbol from the $j$-th DA port with the average power $\mathbb{E}[|x_j|^2]=1$,
and $z_i$ represents the additive white Gaussian noise with variance $\sigma_n^2$ for the $i$-th user.
In this paper, we consider DAS with circular antenna layout as in Figure \ref{figure:sys_model}.
The cell radius is set to $\tilde{R}$ at center $(0,0)$, and the $j$-th DA port is located at
$\left(r \cos{\left(\frac{2\pi (j-1)}{N}\right)}
,r \sin{\left(\frac{2\pi (j-1)}{N}\right)}\right)$ for $j=1, \cdots, N$ with
$r = \sqrt{\frac{3}{7}}\tilde{R}$ as in \cite{Choi:07}.

Here, we assume that each DA port transmits the signal with its full power $P$, or it is turned off.
It was shown in \cite{Anders:98} that such binary on/off power control maintains the optimal performance
for instantaneous channel realizations in two-user environments.
Although the binary power control may not be optimum in terms of the ergodic sum rate
and for the systems with more than two users,
we employ the binary power control for our system to simplify the operations.
Investigation of the optimal power allocation for DAS is outside the scope of this paper,
and remains as an interesting future work.

A main goal of this paper is to determine pairings of $N$ DA ports and their supporting users
which maximize the ergodic sum rate for given user distance information.
Let us denote the transmission mode
\bea
\label{eq:D_definition}
D = [u_1, u_2, \cdots, u_N],
\eea
as the user index of $N$ DA ports where $u_i\in \{0,1,2, \cdots, K\}$ $(i=1,2,\cdots,N)$ represents
the user index supported by the $i$-th DA port $DA_i$.
Here, the index $0$ indicates that no user is supported by the corresponding DA.
Moreover, we define the number of active users and active DA ports as $K_A$ and $N_A$, respectively.
Thus, $K_A$ should be less than or equal to $N_A$, i.e., $K_A\leq N_A \leq N$.
In other words, the supported user indices of $D$ consist of $K_A$ non-zero distinct elements,
and only $N_A$ DA ports have a non-zero user index in $D$.

\section{Ergodic Sum Rate Analysis} \label{sec:Ergodic_SR}
In this section, we will study the statistical properties of the multi-user
multi-DA ports DAS with given user distance information.
The ergodic sum rate $\mathbb{E}[R]$ can be expressed as
\bea
\label{eq:gen_SR}
\mathbb{E}[R] = \sum_{i=1}^{K}{\mathbb{E}[R_i]} = \sum_{i=1}^{K}{\mathbb{E}[\log_2{(1+\rho_i)}]}
\eea
where $\rho_i$ and $R_i=\log_2{(1+\rho_i)}$ indicate the signal to interference plus noise ratio (SINR) and the rate of the $i$-th user, respectively.

From (\ref{eq:D_definition}), let us define $G_i=\{j|u_j=i\}$ as the set of DA port indices supporting the $i$-th user,
$G_T = \bigcup _{i}{G_i}$ as the set of all active DA port indices,
and $G_i^{C} = G_T \backslash G_i$ as the complement of $G_i$ in $G_T$ for $i = 1,2,\cdots, K$ and $j = 1,2,\cdots, N$, respectively.
For user $i$, the signal from DA ports in $G_i$ is regarded as the desired signal,
while the signal transmitted from DA ports in $G_i^{C}$ is treated as interference.
Especially, when $G_i=\emptyset$, which means that user $i$ is not supported by any DA port and thus is not an active user,
the rate for the corresponding user is zero ($R_i=0$).
Note that only $K_A$ active users have non-zero rates.
Then, the rate of the $i$-th active user is generally represented as
\bea
\label{eq:R_i}
R_i&=&\log_2{\left(1+\frac{\rho_{i,S}}{\rho_{i,I}}\right)}\nonumber\\
&=&\log_2{\left(1+\frac{\sum_{k\in G_i}{S_{i,k}P|h_{i,k}|^2}}
{\sigma_n^2+\sum_{l\in G_i^{C}}{S_{i,l}P|h_{i,l}|^2}}\right)}
\eea
where $\rho_{i,S}$ and $\rho_{i,I}$ denote the instantaneous signal power and
the interference plus noise power of user $i$, respectively.

In what follows, we consider the probability density function (pdf) of each user's SINR
to derive a closed form of the ergodic sum rate.
It is obvious from (\ref{eq:R_i}) that
$\rho_{i,S}$ and $\rho_{i,I}$ follow a weighted Chi-squared distribution when $S_{i,j}$ $(i\in\{1, \cdots, K\}$, $j\in\{1, \cdots, N\})$ is fixed.
Thus, the corresponding pdfs can be expressed as
\bea
\label{eq:pdf_numer}
f_{\rho_{i,S}}(\rho)\!=\!\!\sum_{k\in G_i}\!{\frac{1}{S_{i,k}P}\!\!\left(\!\prod_{\substack{l\in G_i \\ l\neq k}}\!
{\frac{S_{i,k}}{S_{i,k}-S_{i,l}}}\!\right)\!\exp{(-\frac{\rho}{S_{i,k}P})}} ~~~~~~~~~~~~~~~\text{for } \rho>0
\eea
and
\bea
\label{eq:pdf_denumer}
f_{\rho_{i,I}}(\rho)\!\!=\!\!\!\sum_{u\in G_i^{C}}\!{\frac{1}{S_{i,u}P}\!\!\left(\!\prod_{\substack{v\in G_i^{C} \\ v\neq u}}\!
{\frac{S_{i,u}}{S_{i,u}\!-\!S_{i,v}}}\!\!\right)\!\exp{(-\frac{\rho\!-\!\sigma_n^2}{S_{i,u}P})}} ~~~~~~~~~~~~~~~\text{for } \rho>\sigma_n^2.
\eea

Applying the Jacobian transformation to (\ref{eq:pdf_numer}) and (\ref{eq:pdf_denumer}),
we obtain the pdf of SINR for user $i$ as
\bea
\label{eq:pdf_SINR}
f_{\rho_{i}}(\rho)\ &=& \int_{\sigma_n^2}^{\infty}{f_{\rho_{i,S}}(\rho \theta) f_{\rho_{i,I}}(\theta) \theta d\theta}\nonumber\\
&=&\frac{1}{P^2}\!\sum_{k\in G_i}\!\sum_{u\in G_i^{C}}\!\!\!\left(\prod_{\substack{l\in G_i \\ l\neq k}}{\frac{S_{i,k}}{S_{i,k}-S_{i,l}}}\right)
\!\!\!\!\left(\prod_{\substack{v\in G_i^{C} \\ v\neq u}}{\frac{S_{i,u}}{S_{i,u}\!-\!S_{i,v}}}\right)\nonumber\\
& &\cdot
{\frac{1}{S_{i,k}S_{i,u}}}\exp{(\frac{\sigma_n^2}{S_{i,u}P})}
\!\!\int_{\sigma_n^2}^{\infty}{\!\exp{(-\frac{(S_{i,u}\rho\!+\! S_{i,k})\theta}{S_{i,k}S_{i,u}P})}}\theta d\theta\nonumber\\
\!\!\!\!\!&=&\!\!\!\!\!\frac{1}{P}\!\sum_{k\in G_i}\!\sum_{u\in G_i^{C}}\!\!\!
\left(\prod_{\substack{l\in G_i \\ l\neq k}}{\frac{S_{i,k}}{S_{i,k}\!-\!S_{i,l}}}\right)
\!\!\!\!\left(\prod_{\substack{v\in G_i^{C} \\ v\neq u}}{\frac{S_{i,u}}{S_{i,u}\!-\!S_{i,v}}}\right)
\!\frac{\sigma_n^2(S_{i,u}\rho\!+\!S_{i,k})\!\!+\!\!S_{i,k}S_{i,u}P}{(S_{i,u}\rho\!+\!S_{i,k})^2}\exp{(-\frac{\sigma_n^2 \rho}{S_{i,k}P})}.
\eea

Then, the ergodic sum rate is derived from (\ref{eq:pdf_SINR}) using integration formulas \cite{TOI:07} as
\bea
\label{eq:ESR}
&\mathbb{E}&\!\!\!\!\![R_i] = \int_{0}^{\infty}{\log_2{(1+\rho)} f_{\rho_i}(\rho)} d\rho \nonumber\\
&=&\!\!\!\frac{1}{\ln{2}}\!\!\sum_{k\in G_i}\!\!\sum_{u\in G_i^{C}}\!\!\!\left(\!\!\prod_{\substack{l\in G_i \\ l\neq k}}\!\!{\frac{S_{i,k}}{S_{i,k}\!\!-\!\!S_{i,l}}}\!\!\right)
\!\!\!\!\left(\!\!\prod_{\substack{v\in G_i^{C} \\ v\neq u}}\!\!{\frac{S_{i,u}}{S_{i,u}\!\!-\!\!S_{i,v}}}\!\!\right)\!\! \frac{S_{i,k}}{S_{i,k}\!\!-\!\!S_{i,u}}\\\nonumber
& &\cdot\!\left\{\!\exp\left(\frac{\sigma_n^2}{S_{i,k}P}\right)\!\!Ei{\left(\frac{\sigma_n^2}{S_{i,k}P}\right)}
\!\!-\!\exp\left(\!\frac{\sigma_n^2}{S_{i,u}P}\!\right)\!\!Ei{\left(\!\frac{\sigma_n^2}{S_{i,u}P}\!\right)}\!\!\right\}
\eea
where $Ei(x) = \int_{x}^{\infty}{\frac{\exp{(-t)}}{t}}dt$ denotes the exponential integral.
Finally, the ergodic sum rate expression for DAS can be obtained
by substituting (\ref{eq:ESR}) into (\ref{eq:gen_SR}).
Note that this is a function of SNR for given user location information.

As a simple example,
we consider the DAS with two users and two DA ports ($N = K = 2$).
First, for the $K_A = 2$ case,
there are two possible transmission modes, i.e., $D = [1, 2]$ and $[2, 1]$.
Similarly, for the single user transmission case ($K_A=1$),
there exist six cases of the transmission mode (i.e. $D = [1, 0]$, $[2, 0]$, $[0, 1]$, $[0, 2]$,$[1, 1]$, $[2, 2]$).
In the former case, for example, $D = [2, 1]$ indicates that user $1$ and user $2$ are supported by $DA_2$ and $DA_1$, respectively,
and the signal transmitted from $DA_1$ is considered as interference to user $1$.
In this case, the supporting DA port index sets for each user are given as
$G_1=\{2\}$, $G_2=\{1\}$, $G_1^{C}=\{1\}$, and $G_2^{C}=\{2\}$.

By plugging (\ref{eq:ESR}) into (\ref{eq:gen_SR}) with this setup,
the ergodic sum rate expression for $D=[2, 1]$ is written as
\bea
\label{eq:ESR_12}
\mathbb{E}[R]&=&\mathbb{E}[R_1]+\mathbb{E}[R_2]\nonumber\\
&=&\frac{1}{\ln{2}}\Bigg[\frac{S_{1,2}}{(S_{1,2}-S_{1,1})}\left\{\exp{\left(\frac{\sigma_n^2}{S_{1,2}P}\right)}Ei{\left(\frac{\sigma_n^2}{S_{1,2}P}\right)}
-\exp{\left(\frac{\sigma_n^2}{S_{1,1}P}\right)}Ei\left(\frac{\sigma_n^2}{S_{1,1}P}\!\right)\right\}\nonumber\\
& &+\frac{S_{2,1}}{(S_{2,1}-S_{2,2})}
\bigg\{\exp{\left(\frac{\sigma_n^2}{S_{2,1}P}\right)}Ei\left(\frac{\sigma_n^2}{S_{2,1}P}\right)-\exp{\left(\frac{\sigma_n^2}{S_{2,2}P}\right)}
Ei\left(\frac{\sigma_n^2}{S_{2,2}P}\right)\bigg\}\Bigg].\nonumber
\eea
In contrast, for the single user transmission case of $D = [1, 1]$, it follows
\bea
\label{eq:ESR_11}
\mathbb{E}[R]&=&\mathbb{E}[R_1]\nonumber\\
&=&\!\!\!\!\!\frac{1}{\ln{2}}\Bigg\{\frac{S_{1,1}}{\left(S_{1,1}-S_{1,2}\right)}\exp{\left(\frac{\sigma_n^2}{S_{1,1}P}\right)}Ei\left(\frac{\sigma_n^2}{S_{1,1}P}\right)
+\frac{S_{1,2}}{\left(S_{1,2}-S_{1,1}\right)}\exp{\left(\frac{\sigma_n^2}{S_{1,2}P}\right)}Ei\left(\frac{\sigma_n^2}{S_{1,2}P}\right)\Bigg\}.\nonumber
\eea

A closed form of the ergodic sum rate for other modes can also be
similarly obtained by using the above derived expressions.
It should be emphasized that the derived ergodic sum rate has a generalized form
with respect to the number of terms in the numerator and the denominator in (\ref{eq:R_i}).
Thus, the derived expression can present the sum rate of DAS with arbitrary numbers of users and DA ports.
It will be shown later in Section \ref{sec:simulation}
that our derived ergodic sum rate expressions accurately match with the simulation results.

\section{Transmission Selection Strategies}
\label{sec:schemes}
In this section, we study the ideal mode selection
and propose a simple transmission selection scheme for the ergodic sum rate maximization
using the derived expression in the previous section.

\subsection{Ideal Mode Selection}
\label{sec:ST}
We first address the ideal mode selection which chooses the optimum transmission mode
by exhaustive search.
To this end, we compute the ergodic sum rates using (\ref{eq:gen_SR}) and (\ref{eq:ESR})
for all possible transmission modes with given user distance information.
Then, we select the best mode which has the highest ergodic sum rate among them.
In this mode selection problem, the pairings of active DA ports and the supported users
become our main consideration to maximize the ergodic sum rate.

Now we examine the number of mode candidates for the ideal mode selection.
First, for the single user transmission case ($K_A = 1$),
it is obvious that supporting one user with all $N$ active DA ports always shows better performance than serving the user with fewer DA ports
due to the increased array gain.
For example, $D=[1, 1, 1]$ generates better sum rate compared to that of $[1,0,1]$ or $[0,0,1]$
because all DA ports are turned on and yield higher signal power.

Generalizing this to arbitrary $N$ and $K$, the size of the set of mode candidates $\mathcal{D}$ is given as $(K+1)^N-K(2^N-2)-1$,
since the single user transmission modes with $N_A = 1,2,\cdots, N-1$ do not need to be included
for the transmission mode selection problem.
It is obvious that the number of mode candidates increases exponentially with $N$,
and thus the search size of the ideal mode selection may become prohibitive as $N$ and $K$ grow large.
Instead, by using the user distance information, we can conceive more efficient transmission methods
which determine the best transmission mode
with reduced mode candidates.


\subsection{Mode Selection based on Minimum Distance}
\label{sec:ST_reduced}
In this subsection,
we propose a transmission selection scheme which reduces the candidate size of the ideal mode selection.
Motivated by the fact that the overall sum rate is determined mostly by the DA port with the nearest user,
we introduce a new method based on the minimum distance where
the number of mode candidates decreases dramatically for large $N$ and $K$.

For the DAS with $K$ users and $N$ DA ports,
we start with the transmission mode where each DA port serves the nearest user from itself with $N_A=N$.
Then, we turn off DA ports one by one with $2^N-1$ distinct combinations,
and generate a mode candidate.
Then, all these $2^N-1$ candidates are added to the mode candidate set $\mathcal{D}$.
As mentioned before, we exclude modes with $N_A=K_A=1$ from $\mathcal{D}$
since those modes have lower rates due to the reduced array gain.
Instead,
we add a mode serving one user with all of $N$ DA ports (i.e. $K_A=1$ and $N_A=N$)
where the only user is chosen as the one
who has the minimum distance among all users and DA ports.
%
%
%
The algorithm is summarized in Table \ref{table:Proposed Algorithm Table}.

Finally, the mode candidate set $\mathcal{D}$ with size of $2^N-N$ is completed for the DAS,
since out of $2^N-1$ DA on/off combinations, we have excluded $N$ modes with $N_A=1$
and added a single user transmission mode with $N_A=N$.
Then, after evaluating the ergodic sum rate for each candidate mode in $\mathcal{D}$
using the expressions derived in Section \ref{sec:Ergodic_SR},
we select the best mode which exhibits the maximum rate.
It is clear that the number of candidates is reduced substantially
compared to that of the ideal mode selection $(K+1)^N-K(2^N-2) -1$.
Note that the number of mode candidates for the proposed scheme is
determined by the number of DA ports, and is independent of the number of users.
As a result, the complexity of our proposed scheme is not affected by $K$,
unlike the ideal mode selection.
Also, as the number of users and DA ports increase,
a reduction in the number of mode candidates in the proposed scheme grows.
For example, for $N=K=5$, $7625$ mode candidates are required for the ideal mode selection,
while $27$ candidates are employed in our proposed selection scheme,
which accounts for only $0.35\%$ of the original search size.

It should be emphasized again that our proposed scheme needs only the user distance information at each DA port
and does not require the instantaneous CSI. Thus, the overhead associated with the CSI feedback can be avoided.
It will be shown in Section \ref{sec:simulation} that
the proposed scheme based on the minimum distance
shows the ergodic sum rate performance identical to the ideal transmission selection scheme.

\section{Analysis on Mode Selection}
\label{sec:Analysis}
In the previous section,
we have proposed a transmission mode selection scheme with substantially reduced number of candidates
using the ergodic sum rate expression derived in Section \ref{sec:Ergodic_SR}.
Normally, different transmission modes exhibit different sum rate curve patterns.
For example, for the DAS with $N=K=2$,
modes with $K_A=N_A=2$ show saturated sum rates at high SNR,
while the performance of the single user transmission mode with $N_A=2$ increases as SNR grows large,
and thus cross-over points exist in the sum rate curves with different modes.
In this section, to provide insightful observations,
we derive the SNR cross-over point of transmission modes
and discuss the mode selection methods in detail.
The derived ergodic sum rate expression in Section \ref{sec:Ergodic_SR} contains a summation of exponential integral forms.
It is difficult to compare the expressions with the sum of the exponential integrals,
and thus the analysis of the mode selection is hard to obtain using exact expressions.
For simple analysis, we compute the cross-over point with the approximated exponential integral
for the system with two DA ports and two users.
In this case,
the number of transmission mode candidates for the ideal mode selection scheme is four in total 
with $\mathcal{D} = \{[1, 1], [2, 2], [1, 2], [2, 1]\}$.

By applying an approximation of the exponential integral $Ei(x)\approx e^{-x}\ln{(1+\frac{1}{x})}$ \cite{EI_app:64}
to the ergodic sum rate expressions of the system with $N=K=2$,
the sum rate for $D=[1,2]$ is written as
\bea
\label{eq:ESRAPP_12}
\mathbb{E}[R] \simeq \frac{1}{\ln{2}}\left\{\frac{S_{1,1}}{(S_{1,2}-S_{1,1})}{\ln{\left(\frac{S_{1,2}\rho+1}{S_{1,1}\rho+1}\right)}}
+\frac{S_{2,2}}{(S_{2,1}-S_{2,2})}\ln{\left(\frac{S_{2,1}\rho+1}{S_{2,2}\rho+1}\right)}\right\}
\eea
where $\rho$ denotes the SNR as $\rho=P/\sigma_n^2$.
Also, the rate for $D=[1,1]$ is written as
\bea
\label{eq:ESRAPP_11}
\mathbb{E}[R] \simeq \frac{1}{\ln{2}}\left\{\frac{S_{1,1}}{(S_{1,1}-S_{1,2})}\ln{(S_{1,1}\rho+1)}
+\frac{S_{1,2}}{(S_{1,2}-S_{1,1})}\ln{(S_{1,2}\rho+1)}\right\}.
\eea
The approximated expressions for $D= [2,1]$ can be derived similarly.

To determine the cross-over point of the ergodic sum rate between $D = [1,1]$ and $[1,2]$,
we first let (\ref{eq:ESRAPP_12}) and (\ref{eq:ESRAPP_11}) equal.
Then, it follows
\bea
\label{eq:Cross-over_derivation1}
\ln{(1+S_{1,2}\rho)} = \frac{S_{2,2}}{(S_{2,1}-S_{2,2})} \ln\left(\frac{1+S_{2,1}\rho}{1+S_{2,2}\rho}\right).
\eea
Applying high SNR assumptions to both sides of (\ref{eq:Cross-over_derivation1}) as
\bea
\label{eq:Cross-over_derivation2}
\ln{(S_{1,2}\rho)} = {\frac{S_{2,2}}{(S_{2,1}-S_{2,2})}}\ln{\left(\frac{S_{2,1}}{S_{2,2}}\right)},
\eea
we notice that equation (\ref{eq:Cross-over_derivation2}) has only one solution with respect to $\rho$.
Finally, the SNR cross-over point between transmission modes, $\rho_c$, is computed using (\ref{eq:Cross-over_derivation2}) as
\bea
\label{eq:Cross-over}
\rho_c = \frac{1}{S_{1,2}}\left(\frac{S_{2,1}}{S_{2,2}}\right)^{\frac{S_{2,2}}{S_{2,1}-S_{2,2}}}.
\eea

Thus, the two user transmission mode $D=[1,2]$ is selected if the SNR is smaller than $\rho_c$,
while the single user transmission mode $D=[1,1]$
is chosen at SNR higher than $\rho_c$.
The cross-over point between $[1,1]$ and $[2,1]$ can be similarly computed as
$\rho_c= \frac{1}{S_{2,1}}(\frac{S_{2,1}}{S_{2,2}})^{\frac{S_{2,1}}{S_{2,1}-S_{2,2}}}$.
This indicates that a single user transmission mode is expected to exhibit
better performance compared to a two user transmission mode at high SNR region.
The accuracy of our analysis will be verified numerically through the Monte Carlo simulations
in Section \ref{sec:simulation}.

Moreover,
since the logarithm increases monotonically with its argument, a sum rate expression for $D=[i,i]$ in (\ref{eq:ESRAPP_11})
is lower-bounded by
\bea
\label{eq:approximate_bound}
\mathbb{E}[R] & \geq & \frac{S_{i,1}\ln{\left(\text{max}(S_{i,1},S_{i,2})\rho+1\right)-S_{i,2}\ln{\left(\text{max}(S_{i,1},S_{i,2})\rho+1\right)}}}
{(S_{i,1}-S_{i,2})\ln{2}}\nonumber\\
& = & \log_2{\left(\text{max}(S_{i,1},S_{i,2})\rho+1\right)}~~~~~~~~~~~~~~~~~~~~~~~~~~~~~~~~~~~~~~~~\text{for $i={1,2}$.}
\eea
Since the minimum distance user maximizes $\text{max}(S_{i,1},S_{i,2})$ for the single user transmission,
this indicates that a user should be determined based on the minimum distance between users and DA ports.

\section{Simulation Results} \label{sec:simulation}
In this section, we will confirm the accuracy of the derived ergodic sum rate expression and
present the performance of the proposed mode selection scheme via Monte Carlo simulations.
Throughout the simulation, the pathloss exponent $p$ and the cell radius $\tilde{R}$ are set to be $3$ and $\sqrt{\frac{112}{3}}$,
respectively.\footnote{
With this setting, the cell edge users have a received SNR loss of $23.5~\text{dB}$ compared to the cell center
for the conventional pathloss modeling.}
The number of generated channel realizations is equal to $5000$.

\begin{figure}
\begin{center}
\includegraphics[width=5.5in]{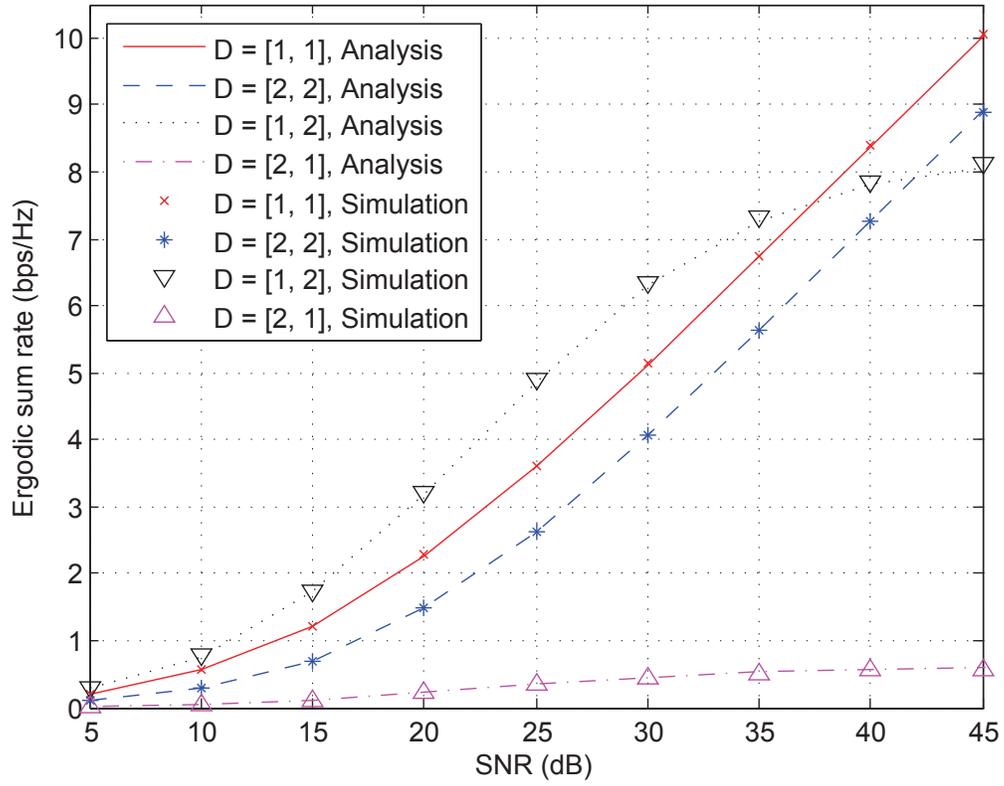}
\end{center}
\caption{Ergodic sum rates for two users and two DA ports
where two users are located at ($-3, -2.5$) and ($3, 3.5$)}
\label{figure:confirm_derivation}
\end{figure}

In Figure \ref{figure:confirm_derivation}, we illustrate the ergodic sum rate of each transmission mode
for the DAS with two users and two DA ports with fixed user locations.
The location of two DA ports are set to $(4,0)$ and $(-4,0)$,
and it is assumed that two users are located at ($-3, -2.5$) and ($3, 3.5$).
From this figure,
we can verify that our derived ergodic sum rate expressions accurately match with the simulation results.
We can check that there exist cross-over points among different modes,
and the best mode for maximizing the ergodic sum rate varies according to SNR.
Using the analysis results in Section \ref{sec:Analysis},
the SNR cross-over point between two modes $D=[1,1]$ and $[1,2]$ is computed as $\rho_c=37.2 ~\text{dB}$,
and this shows a good agreement with the simulation results in Figure \ref{figure:confirm_derivation}.
%
%
%
%
Furthermore, the sum rate curves for each mode exhibit different trends depending on the SNR.
For instance,
$D=[1,2]$ is the best mode to maximize the ergodic sum rate in low and mid SNR region,
while $D=[1,1]$ becomes the optimal mode in high SNR region over $\rho_c=37.2~\text{dB}$.
This emphasizes the importance of transmission mode selection
to maximize the overall performance in the multi-user multi-DA port scenario.

\begin{figure}
\begin{center}
\includegraphics[width=5.5in]{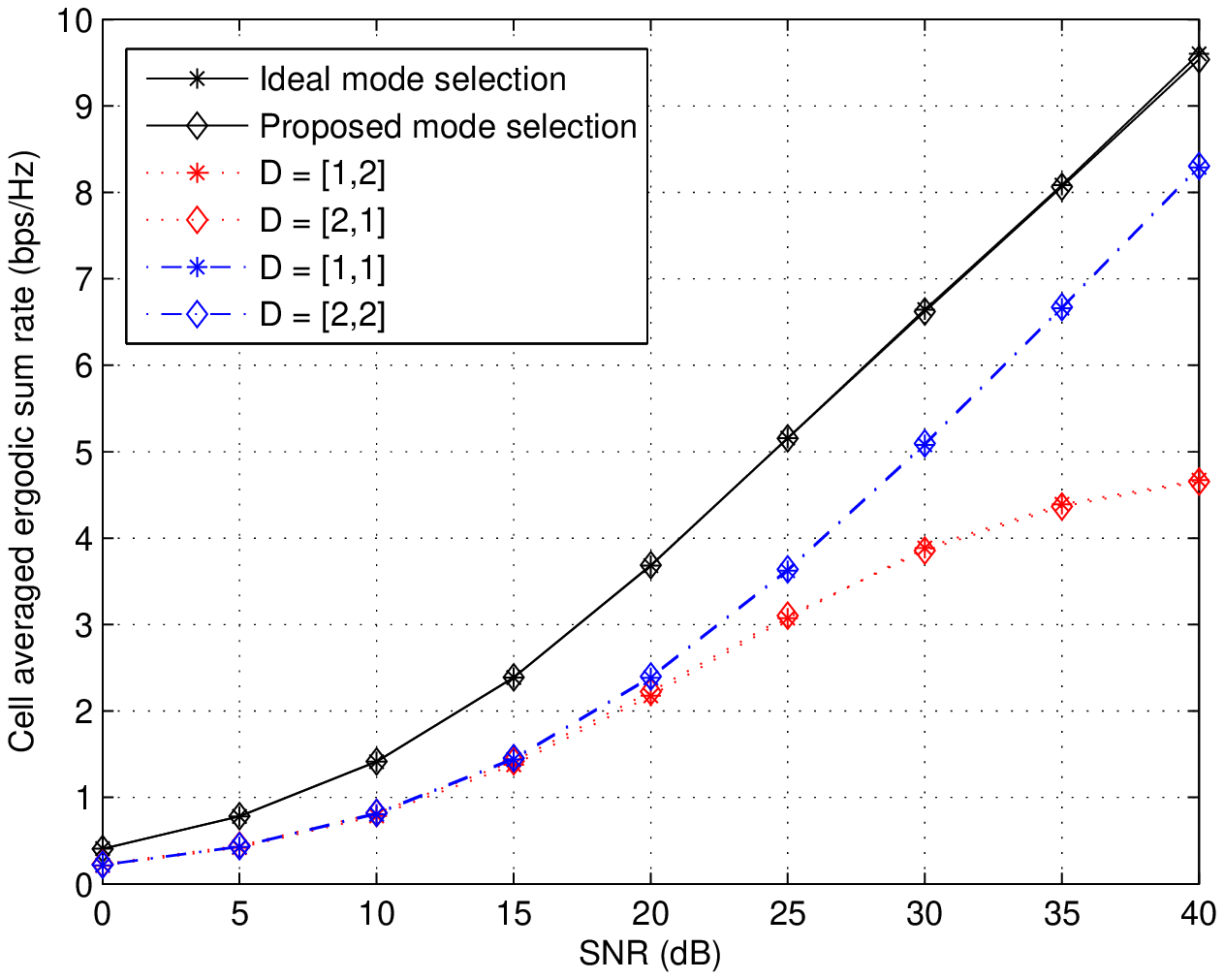}
\end{center}
\caption{Average ergodic sum rate for DAS with $N = K = 2$}
\label{figure:two_user}
\end{figure}

Next, Figure \ref{figure:two_user} presents the simulation results employing the proposed transmission selection scheme
with random user locations.
Users are randomly generated with a uniform distribution within a cell
and the number of user generations is set to $4000$.
Figure \ref{figure:two_user} evaluates the cell averaged ergodic sum rate performance
of the proposed transmission selection scheme with $N=K=2$.
To confirm the performance of our proposed scheme, we also plot the ergodic sum rate performance of the ideal mode selection
where the sum rates are calculated by averaging actual channel realizations for all possible transmission modes.
We can notice that the performance of our proposed scheme based on the minimum distance
is identical to that of the ideal mode selection.
The sum rate performance with each mode is also presented.
Here, $D=[1,2]$ and $[2,1]$, and $D=[1,1]$ and $[2,2]$ exhibit the same performance, respectively,
because the sum rates for all user positions are averaged in a cell.
The two user transmission modes ($K_A=2$) show no difference in the averaged sum rate compared with the single user transmission modes ($K_A=1$)
for low SNR, but the saturated performance is observed at high SNR.
This is due to a fact that the other users' interference power degrades the sum rate performance severely for high SNR
where interference becomes a dominant factor.
It is obvious from the plot that the proposed mode selection scheme exploits a selection gain
over fixed transmission modes.

\begin{figure}
\begin{center}
\includegraphics[width=5.5in]{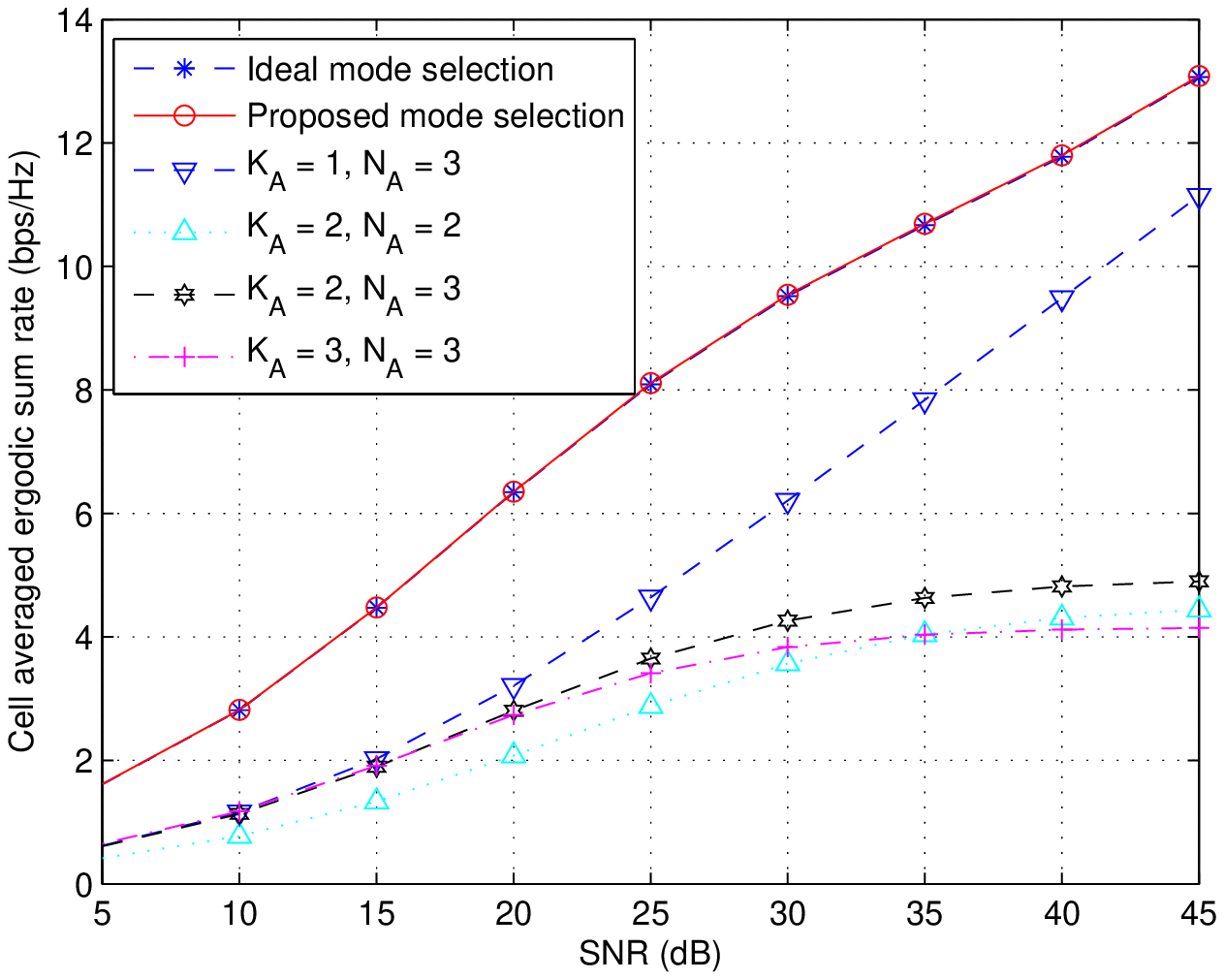}
\end{center}
\caption{Average ergodic sum rate for DAS with $N = K = 3$}
\label{figure:three_user}
\end{figure}

Figure \ref{figure:three_user} presents the cell averaged ergodic sum rate performance
in the DAS with $N=K=3$.
Although the proposed scheme chooses the best mode with reduced mode candidates,
the obtained ergodic sum rate performance is the same as
the ideal mode selection.
In this figure, similar to Figure \ref{figure:two_user}, the cell averaged ergodic sum rate performance of
the single user transmission modes approaches that of the ideal mode selection in high SNR region,
since single user transmission outperforms multi-user transmission in interference limited environments.

\begin{figure}
\begin{center}
\includegraphics[width=5.5in]{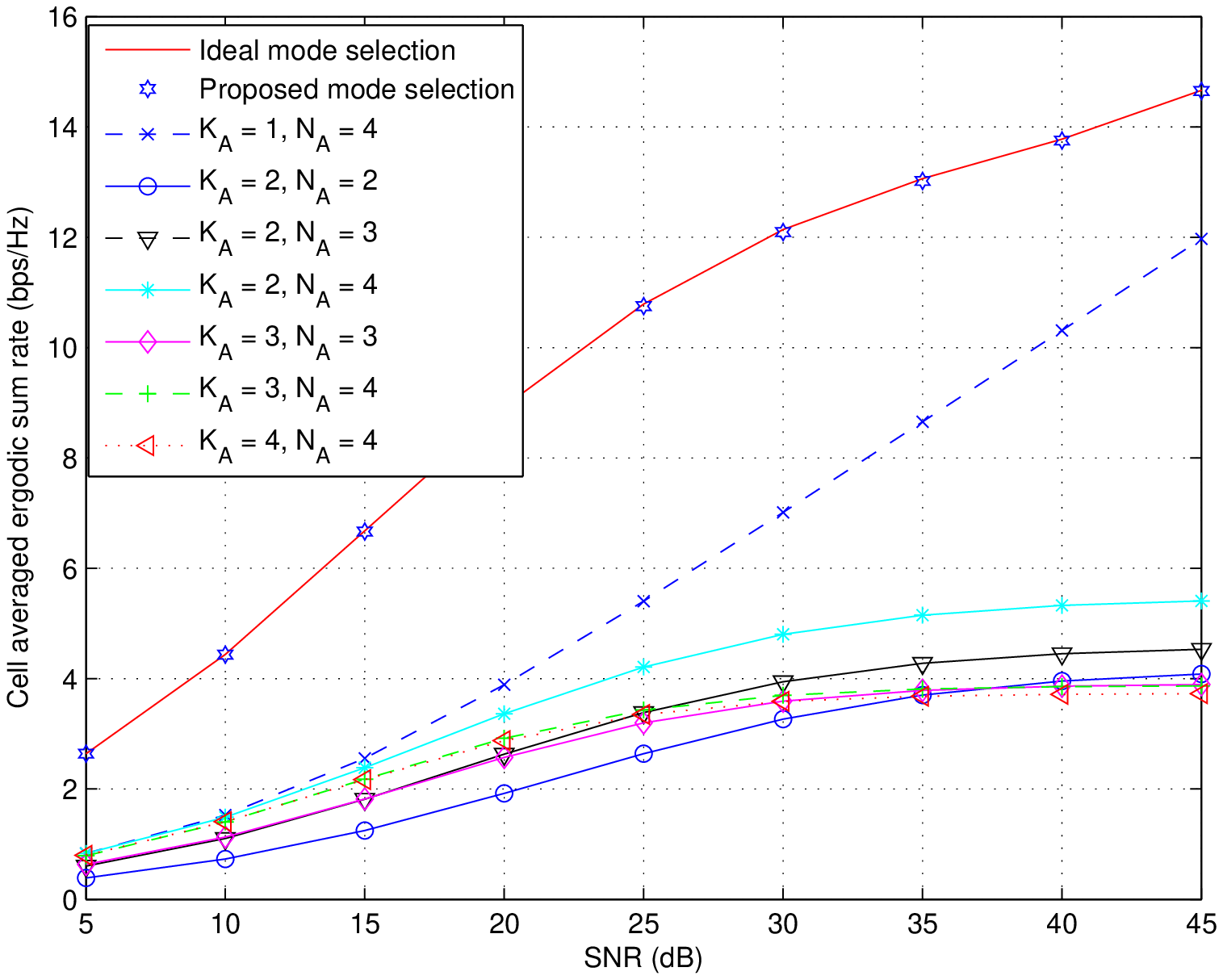}
\end{center}
\caption{Average ergodic sum rate for DAS with $N = K = 4$}
\label{figure:four_user}
\end{figure}

In Figure \ref{figure:four_user}, we plot the performance of the DAS with $N=K=4$.
Similar to the previous cases, in high SNR region, the single user transmission modes
exhibit the best sum rate compared to other modes,
while the curves for the multi-user transmission modes are flattened.
Here, the performance of modes for the multi-user case is saturated with a higher ergodic sum rate value
as the number of active users $K_A$ decreases and the number of active DA ports $N_A$ increases.
It should be emphasized again that our proposed scheme based on the minimum distance reduces the number of mode candidates substantially
without any performance loss compared to the ideal mode selection.
The number of mode candidates for the ideal mode selection is $568$ for $N=K=4$,
while the proposed scheme requires only $12$ mode candidates.
It is clear that savings on the candidate set size reduction becomes significant as the number of users and DA ports increases.

\begin{figure}
\begin{center}
\includegraphics[width=5.5in]{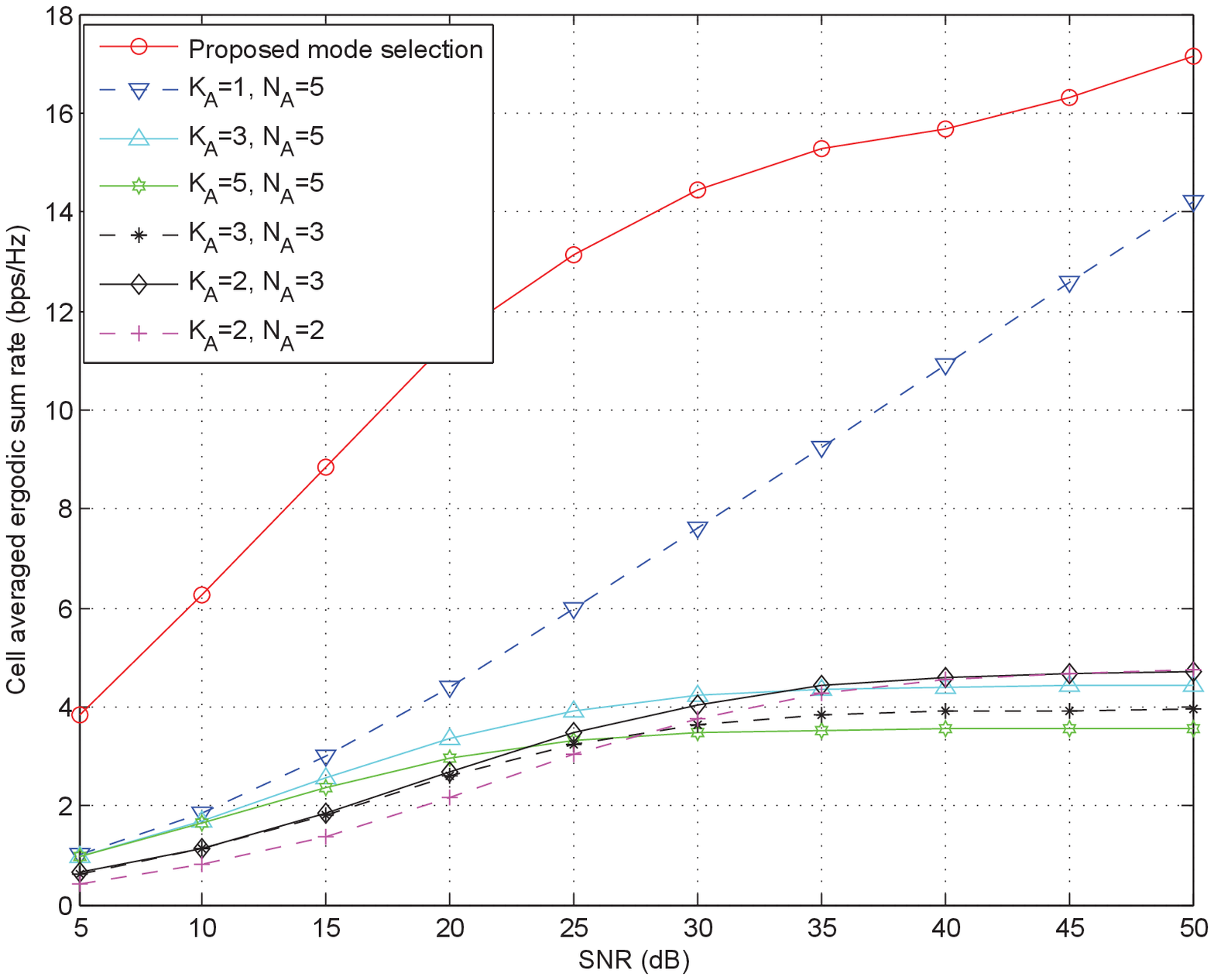}
\end{center}
\caption{Average ergodic sum rate for DAS with $N = K = 5$}
\label{figure:five_user}
\end{figure}

Figure \ref{figure:five_user} presents the average ergodic sum rate for DAS with $N=K=5$.
In this case, the results for the ideal mode selection are not included,
since the number of candidates is $7625$ and the simulations for this case become prohibitive.
In this plot, we observe similar trends as before.
By comparing Figures \ref{figure:three_user}, \ref{figure:four_user}, and \ref{figure:five_user},
we can see that an ergodic sum rate gain obtained by the proposed mode selection method
grows as $N$ and $K$ become large compared to the individual fixed transmission modes.
%
In addition, a transmission mode selection gain decreases as the SNR becomes large.

\begin{figure}
\begin{center}
\includegraphics[width=5.5in]{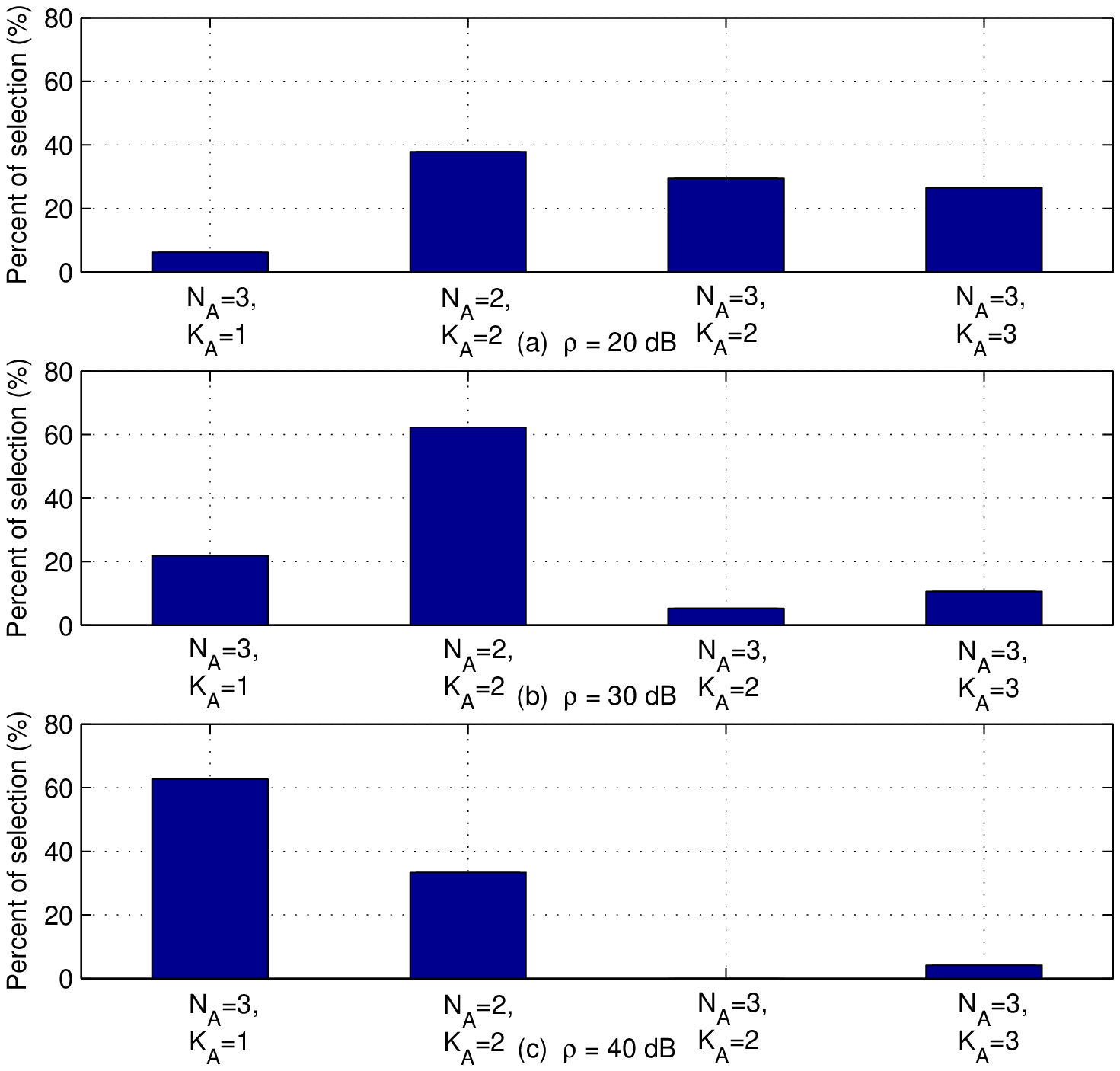}
\end{center}
\caption{Relative occurrences of the selected transmission modes for DAS with $N = K = 3$}
\label{figure:pdf_three}
\end{figure}

In Figure \ref{figure:pdf_three},
we plot the histogram of the selected transmission modes in various SNR ranges
and categorize the transmission modes into $4$ groups according to $K_A$ and $N_A$
for DAS with $N=K=3$.
From Figure \ref{figure:pdf_three},
it is obvious that the single user transmission modes are mostly selected as the best mode
as the SNR increases in DAS.
Thus, it is expected that a gain obtained by the proposed mode selection is reduced at high SNR region
over the performance of modes with $N_A=N$ and $K_A=1$, which have the best sum rate performance among all modes.
This can also be seen as a reason why the sum rate performance of single users transmission modes approaches
that with mode selection as shown in Figures \ref{figure:two_user} $-$ \ref{figure:five_user}.

\begin{figure}
\begin{center}
\includegraphics[width=5.5in]{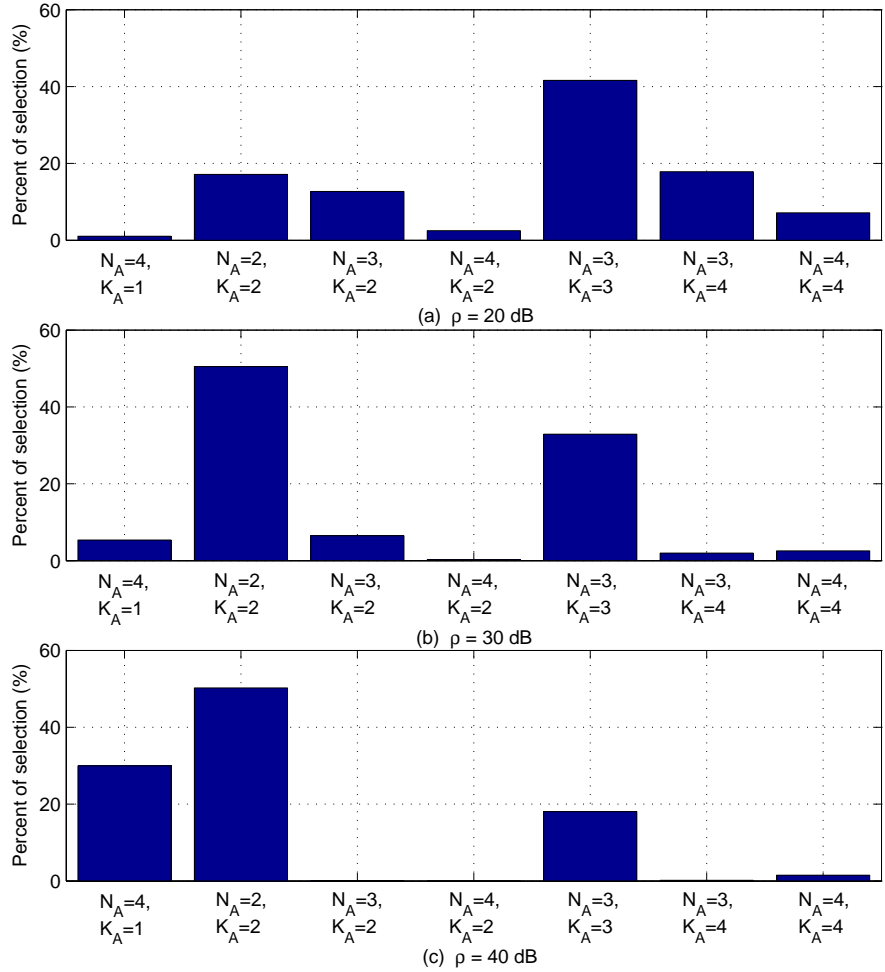}
\end{center}
\caption{Relative occurrences of the selected transmission modes for DAS with $N = K = 4$}
\label{figure:pdf_four}
\end{figure}

Similarly, we illustrate the histogram of selected modes for the DAS with $N=K=4$ in Figure \ref{figure:pdf_four}.
Compared to the DAS with $N=K=3$ in Figure \ref{figure:pdf_three},
the probability that the single user transmission modes are selected is reduced for $N=K=4$.
%
Therefore, although single user transmission modes show better sum rate performance than multi-user transmission modes,
a gain obtained by applying the mode selection method over the modes with $K_A=1$ is expected to grow
with high probability as $N$ and $K$ increase.
We can observe this characteristics in Figures \ref{figure:two_user}, \ref{figure:three_user} and \ref{figure:four_user}.
In addition, since interference becomes a dominant factor for sum rate performance as the SNR increases,
multi-user modes with fewer active DA ports are preferred as shown in Figures \ref{figure:pdf_three} and \ref{figure:pdf_four}.


\section{Conclusions} \label{sec:conclusion}
In this paper, we have studied the multi-user multi-DA port downlink DAS
and have derived an ergodic sum rate expression using the pdf of users' SINR.
Based on the derived expressions,
we have proposed a transmission selection scheme to maximize the ergodic sum rate.
In the proposed scheme, mode candidates are generated by considering
pairings of each active DA port and the nearest user.
Then, we select the mode which maximizes the ergodic sum rate.

The number of mode candidates is reduced dramatically without any performance loss by applying the proposed scheme.
Also, we have analyzed the SNR cross-over points for different transmission modes.
The effectiveness of the proposed scheme has been confirmed through simulations.
We have verified the sum rate performance of the proposed scheme
for various configurations.

\bibliographystyle{ieeetr}
\input{bibliography.filelist}

\clearpage
\newpage
\clearpage
\newpage
\clearpage
\newpage
\clearpage
\newpage
\clearpage

\newpage
\clearpage

\newpage
\clearpage

\newpage
\clearpage
\newpage
\clearpage
\begin{table}[h]
\caption{{Mode Selection based on minimum distance}}
\centering
\begin{tabular}{l}
\hline
${\text{Set}}$ ${D_0=[u_1,\cdots,u_N]}$ ${\text{as the mode}}$ ${\text{where each DA port serves}}$
${\text{its nearest user}}$${\text{ with}}$ ${N_A=N.}$\\
${\text{Initialize } \mathcal{D} = \{D_0\}.}$\\
${\textbf{for }m=1:2^N-1 }$\\
$~~~~~{\text{Generate an }}$${N}$${\text{-digit binary number}}$ ${b = \text{binary}(m)},$ ${\text{where each bit}}$
${\text{represents}}$ ${\text{on/off for the}}$\\ $~~~~~\text{{corresponding DA port.}}$\\
$~~~~~{\textbf{for }i=1:N}$\\
$~~~~~~~~~~~~{\widetilde{u_{i}} \leftarrow 0}~~~~~~~$${\text{if }b(i)=0}$
$~{\text{where} ~b(i)}$ ${\text{denotes the }i\text{-th bit of }b.}$\\
$~~~~~~~~~~~~{\widetilde{u_{i}} \leftarrow u_{i}}~~~~~~$${\text{if }b(i)=1}$\\
$~~~~~{\textbf{end }}$\\
$~~~~~{\text{Add }}$${[\widetilde{u_{1}},\cdots,\widetilde{u_{N}}]}$${\text{ to }\mathcal{D}}$
$~~~~~{\text{if}}$ ${\text{the number of non-zero elements }}$${\text{in the mode is greater than } 1.}$\\
${\textbf{end }}$\\
${\text{Set~} (i^*,j^*) = \arg\min_{i}\min_{j} {d_{i,j}}}$.\\
${\text{Add }}$$\small{[i^*,\cdots,i^*]}$$\small{\text{ to }\mathcal{D}.}$\\
\hline
\end{tabular}
\label{table:Proposed Algorithm Table}
\end{table}
\newpage
\end{document}